# Binding SNOMED CT Terms to Archetype Elements: Establishing a Baseline of Results

I. Berges[1], J. Bermúdez, A. Illarramendi
University of the Basque Country, Donostia – San Sebastián, Spain.

**Summary**

*Background:* The proliferation of archetypes as a means to represent information of Electronic Health Records has raised the need of binding terminological codes -such as SNOMED CT codes- to their elements, in order to identify them univocally. However, the large size of the terminologies makes it difficult to perform this task manually.

*Objectives:* To establish a baseline of results for the aforementioned problem by using off-the-shelf string comparison-based techniques against which results from more complex techniques could be evaluated.

*Methods*: Nine Typed Comparison Methods were evaluated for binding using a set of 487 archetype elements. Their recall was calculated and Friedman and Nemenyi tests were applied in order to assess whether any of the methods outperformed the others.

*Results*: Using the qGrams method along with the 'Text' information piece of archetype elements outperforms the other methods if a level of confidence of 90% is considered. A recall of 25.26% is obtained if just one SNOMED CT term is retrieved for each archetype element. This recall rises to 50.51% and 75.56% if 10 and 100 elements are retrieved respectively, that being a reduction of more than 99.99% on the SNOMED CT code set.

*Conclusions:* The baseline has been established following the above-mentioned results. Moreover, it has been observed that although string comparison-based methods do not outperform more sophisticated techniques, they still can be an alternative for providing a reduced set of candidate terms for each archetype element from which the ultimate term can be chosen later in the more-than-likely manual supervision task.

**Keywords:** Archetype, SNOMED CT

---

[1] **Corresponding author:**
 Idoia Berges.
Computer Science Faculty, University of the Basque Country.
Paseo Manuel de Lardizabal, 1
20018 Donostia – San Sebastián, Spain
Phone: (+34) 943 018151
e-mail: idoia.berges@ehu.es

## 1. Introduction

During the past few years a great effort has been made in order to achieve semantic interoperability of Electronic Health Records (EHRs)[1,2]. Among other benefits, seamless access, any time, any place, to the clinical data of a patient –regardless of where they were recorded- can prevent from deaths and other serious consequences that usually derive from the lack of instant access to vital information such as adverse drug reaction histories. Moreover, interoperability can eliminate necessary duplication of tests and facilitate collaboration between many professionals at distinct points of care [3]. This leads to a safer and cheaper healthcare.

However EHR interoperability is not an easy task, due to the heterogeneity of EHRs. In order to tackle this problem, several standards –such as openEHR[4], CEN 13606[5] and HL7 CDA[6]- have arisen. All of them rely on the idea of defining archetypes for representing knowledge elements. For example, the Blood Pressure archetype of openEHR indicates that a blood pressure record may contain information about the systolic and the diastolic blood pressures. For each of these elements within an archetype, a "text" property indicating their name and a "description" is usually included. Moreover, the elements of the archetypes can be bound to codes of well-known terminologies such as SNOMED CT[7] in order to specify their meaning univocally and increase greatly the chances of interoperability. Unfortunately, the thousands of terms in SNOMED CT make it hard for physicians to find the appropriate bindings, and it results in most cases in a large number of unbounded elements within the archetypes.

For this reason, several approaches have been presented in order to perform bindings automatically. Authors in [8] present a method based on the TF-IDF measure. The work on [9] relies on the UMLS[10] thesaurus as middleman. Moreover, the work on [11] combines linguistic techniques along with information retrieval and context-based techniques. Finally in [12] lexical, linguistic and context-based techniques are used. The recall obtained by these approaches ranges from 55% to 71%, as stated in [12]. All of the solutions are designed using a composition or sequence of elaborated methods, and although we acknowledge their results, we think that in order to be able to fully assess their goodness, a prior assessment of simpler approaches (i.e. simpler from the point of view of their design) is needed. Moreover, we believe that in general, getting almost effortlessly a small set of candidates for each binding suffices, since code binding is a sensitive task that will require some sort of human supervision.

## 2. Objectives

The main goal of this paper is to establish a baseline of results for the problem of binding SNOMED CT codes to archetype elements. The SNOMED CT FSN descriptions from the January 2010 release are used for that purpose. This baseline will be established from the results obtained by syntax-based string comparison methods. These methods are usually seen as naive but will enable a more accurate assessment of the results of other more sophisticated techniques (e.g. based on semantics), which presumably will outperform them but require an elaborate design. In other words, the baseline will help to assess how much of the goodness of results of the sophisticated methods could also be achieved by simpler, off-the-shelf methods. Three sub-objectives arise from the main goal:
- Calculation of the recall of several string comparison methods.
- Statistical comparison of the results obtained by these methods in order to determine whether any of them outperforms the others significantly.
- Drawing conclusions from the two previous points in order to establish the aforementioned baseline.

## 3. Methods

### 3.1 Dataset

487 archetype elements were gathered from the set of 25 archetypes with manual SNOMED CT bindings found in [13]. The choice of this dataset was due to the large amount of bindings it contains in comparison with other datasets we were familiar with. The root node of all the archetypes had been mapped to SNOMED CT Observable entities. Two pieces of information were selected from each archetype element: "text" and "description". Moreover, we created a third piece of information, called "extended text", which combines the 'text' with the name of the root node of the source archetype. One example of archetype element is the following one, extracted from the archetype whose root node is *Uterine contractions*:
- Text : Very strong
- Description: Possibly excessively strong contractions
- Extended text: Uterine contractions very strong
- SNOMED CT binding: 289720004 (Very strong uterine contractions (finding))

It must be pointed out that not every element contains a relevant 'description' field, so in the end we were able to retrieve 487 texts, 254 descriptions and 487 extended texts.

### 3.2 Comparison methods

Three different syntactic comparison methods were chosen for our evaluation: Levenshtein distance [14], Jaccard similarity [15] and qGrams comparison [16]. Each of them is based on one different approach (character-based, word-based, and n-gram-based, respectively)

**Levenshtein distance:** This method takes into account the minimum number of insertions (*Ins*), deletions (*Del*) and substitutions (*Subs*) of single characters that are needed in order to transform one string into the other. The distance can be normalized by dividing it by the length of the longest string. For example, five edit operations are needed in order to transform the string "rimmel" into "spine" (*Subs*(r,s) -> "simmel"; *Ins*(p) -> "spimmel; *Del*(m) -> "spimel"; *Subs*(m,n) -> "spinel"; *Del*(l) -> "spine")

**Jaccard similarity:** This method divides the strings $S_A$ and $S_B$ into their corresponding set of words A and B. Then, the similarity index is calculated as:

$$J(S_A, S_B) = \frac{|A \cap B|}{|A \cup B|}$$

For instance, the Jaccard similarity of strings "dry skin" and "colour of the skin" is calculated as follows:
|{skin}| / {dry,skin,colour,of,the}|= 1/5 = 0.2

**qGrams comparison:** In qGrams each of the strings to be compared is divided into a set of n-grams, that is, substrings of length 'n', with n=3 in our case. For instance, the string "heart rate" would be converted to the set [##h, #he, hea, ear, art, rt#, t#r, #ra, rat, ate, te#, e##]. Then, the similarity index between both sets is calculated in the same way as explained above for the Jaccard similarity.

Given that we consider three types of information (texts, descriptions, extended texts) and three comparison methods (Levenstein, Jaccard and qGrams), nine different <Type of information, Comparison method> pairs can be formed. We call TCM (Typed Comparison Method) to each of those pairs. That is, in order to calculate the binding for each of the 487 archetype elements, one type of information and one comparison method was used at a time. The archetype element was compared against the set of 388289 terms in SNOMED CT. The result of that comparison was the list of the 388289 terms ordered by similarity to the archetype element. The process was repeated for each TCM (for the archetype elements lacking description, the corresponding TCMs were excluded).

### 3.3 Statistical measures and tests

Several statistical measures and tests were used for evaluation. First, the recall of each TCM was calculated. Then, the Friedman test [17] was applied to find out whether the choice of TCM has any effect in the result of the comparisons. Finally, the Nemenyi test [18] allowed evaluating whether noticeable differences exists between the performances of two TCMs. The machine learning community has acknowledged these Friedman plus Nemenyi tests as reference and globally accepted statistical tools to assess the significance of the differences when several techniques are compared in different scenarios [18,19]. An explanation of how these measures and tests were used is given below.

**Recall:** We calculated the recall $r_T^{TCM}$ of each TCM in order to evaluate its performance (See results in section 4.1)

$$r_T^{TCM} = \frac{\#correct\ bindings}{\#total\ bindings}$$

where a binding is considered *correct* if the manually selected term is retrieved among the top T terms on the list retrieved by the TCM.

**Friedman test:** This non-parametric test was applied in order to detect whether there were significant differences among the performance of the nine TCMs (See results in section 4.2). In this case only the archetype elements which include a description were taken into account. The comparison results are stored in a table of *n* rows and *k* columns, being *n* the number of elements and *k* the number of TCMs. In our case it yields a 254x9 table. Each cell contains the number of SNOMED CT terms whose similarity value with the archetype element considered in that row, computed using the TCM considered in that column, is equal or greater than the similarity value of the SNOMED CT term that was indicated as correct result in the manually bounded archetype elements (That is, how many SNOMED CT terms are considered better binding candidates by the TCM than the actual correct result). Then, the values of each cell are substituted for their rank order inside their row. As in other statistical tests, the null hypothesis $H_0$ is stated as follows:

- Hypothesis $H_0$: The TCM has no effect in the result of the comparison
- Reject $H_0$ if: $F \geq$ critical value at $(1-\alpha)$ in $\chi^2$ distribution table with k-1 degrees of freedom, being $(1-\alpha)$ the confidence level we want to achieve.
- F is calculated as:

$$F = \frac{12}{nk(k+1)} \sum_{j=1}^{k} R_j^2 - 3n(k+1)$$

where $R_j$ = sum of the ranks in column $j$

The test was performed for both α = 0.05 and α = 0.10. Results can be found in section 4.2

**Nemenyi test:** When the Friedman test rejects the null hypothesis $H_0$, Nemenyi's post hoc test can be applied to detect which columns (in this case, which TCMs) are significantly different from each other in terms of their performance. This was necessary in our case, since as it will be seen later, the aforementioned hypothesis $H_0$ was rejected for both values of α. The test compares columns in pairs. A new hypothesis $H'_0$ is stated:

- Hypothesis $H'_0$: The performance of TCMs *i* and *j* is not significantly different
- Reject $H'_0$ if the difference in their corresponding average ranks (*Diff*) is at least the critical difference (CD):

$$Diff(i, j) \geq CD$$ where:

$$Diff(i, j) = \left| \frac{R_i}{n} - \frac{R_j}{n} \right|$$

$$CD = q_\alpha \sqrt{\frac{k(k+1)}{12n}}$$

$q_\alpha$ = value at α in the Studentised range statistic table with ∞ degrees of freedom.

The test was performed for both α = 0.05 and α = 0.10. Results can be found in section 4.3.

## 4. Results

### 4.1 Recall

Fig. 1 shows the recall for each of the TCMs when sets of top terms of different size are considered.

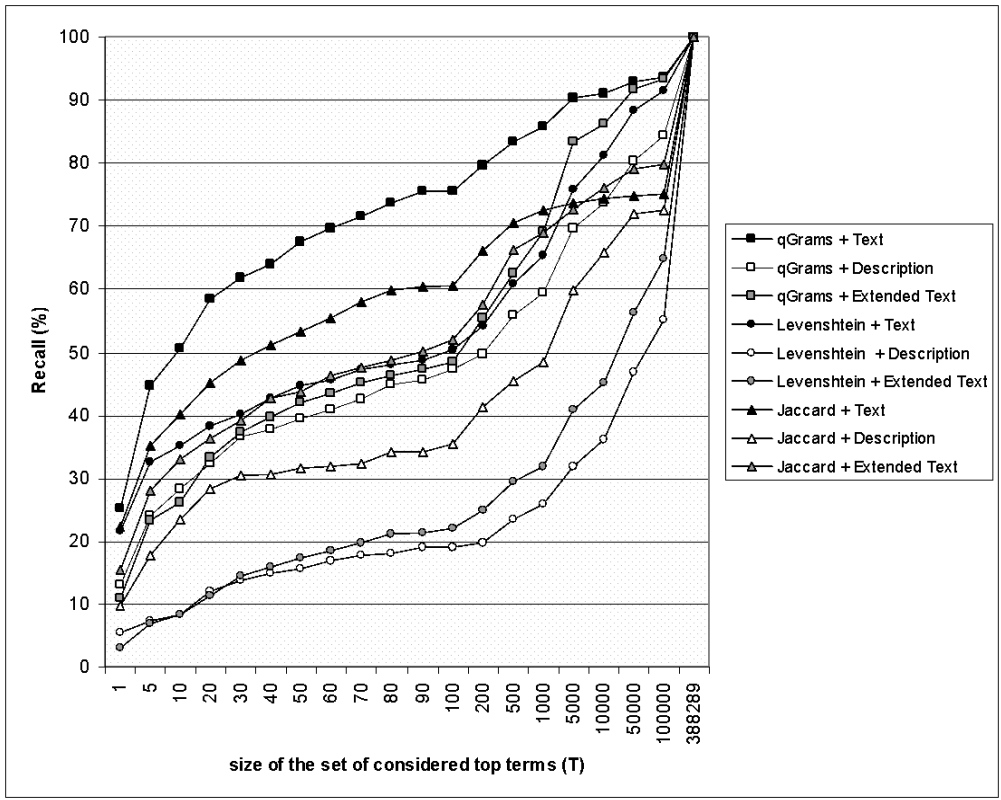

**Fig. 1 Recall of the different TCMs**

It can be seen that using qGrams along with the Text piece of information outperforms the rest of TCMs for most set sizes. Thus, it seems that the results obtained by the TCM qGrams + Text could be taken as a baseline of results for the problem of binding SNOMED CT codes to archetype elements. However, in order to sustain our perception, we wanted to see whether the differences between TCMs were significantly different. As a result, Friedman test was applied.

### 4.2 Results of Friedman test

Table 1 summarizes the values used in order to calculate the Friedman value F.

| Ranks | | | Test Statistics | |
|---|---|---|---|---|
| TCM | $R_{TCM}$ | Mean | | |
| qGrams + Text | 839 | 3.303 | N° of elements (n) | 254 |
| qGrams + Description | 1270 | 5 | | |
| qGrams + Extended T. | 1015.5 | 3.998 | N° of TCMs (k) | 9 |
| Levenshtein + Text | 1202.5 | 4.734 | $\chi^2$ critical value at α=0.05 with 8 degrees of freedom | 15.507 |
| Levenshtein + Description | 1777.5 | 6.998 | | |
| Levenshtein + Extended T. | 1582.5 | 6.23 | $\chi^2$ critical value at α=0.10 with 8 degrees of freedom | 13.362 |
| Jaccard + Text | 1059 | 4.169 | | |
| Jaccard + Description | 1509.5 | 5.943 | F | 378.64 |
| Jaccard + Extended T. | 1174.5 | 4.624 | | |

**Table 1 Relevant values for Friedman test**

Hypothesis $H_0$ stated that the TCM has no effect in the result of the comparison. The hypothesis is rejected if F ≥ $\chi^2$ critical value at α with k-1 degrees of freedom. In this case hypothesis $H_0$ is rejected for both α=0.05 and α=0.10, since 387.64 ≥ 15.507 and 387.64 ≥ 13.362 respectively. As expected, Friedman test concludes that significant differences exist between the TCMs. Then, a post hoc test such as Nemenyi can be used to determine which TCMs differ significantly from the others.

### 4.3 Results of Nemenyi test

Fig.2a shows the results of the pair-wise comparison of TCMs with the Nemenyi test for α = 0.05.

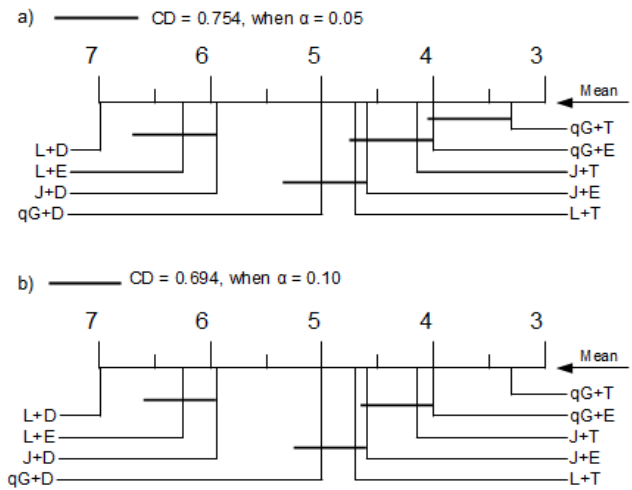

**Fig. 2 Pair-wise comparison of TCMs**

The value $q_\alpha$ in the Studentised range statistic table with ∞ degrees of freedom for this level of confidence is 4.387, which after substituting the corresponding variables in the formula for the Nemenyi test presented in section 3.3, yields a critical difference CD = 0.754. Mean ranking for each TCM (see Table 1) is indicated in the figure. Hypothesis $H_0$ is rejected for those pairs whose difference in mean ranks is at least the critical difference. Thus, two TCMs that are connected implies that there is not enough evidence to assume that their performance differs significantly. For example, at this level of confidence (95%, see Fig. 2a) it can not be concluded that the TCM qGrams + Text outperforms the TCM qGrams + Extended Text.

However, if the level of confidence is lowered to 90% (that is, if α = 0.10, see Fig.2b), we get that $q_\alpha$ = 4.037 and CD = 0,694. In that case some of the previous connections no longer apply. The new scenario shows that at this level of confidence there is enough evidence to conclude that TCM qGrams + Text outperforms the TCM qGrams + Extended Text (as well as the remaining TCMs).

5. Discussion and conclusions

Binding SNOMED CT terms to archetype elements is a very important task on the way to EHR interoperability. For that reason, methods that facilitate this task are needed. As it was expected, results show that the considered syntactic-based string comparison methods do not outperform other approaches (see Section 1). Since we never aimed at obtaining better results than those sophisticated methods, we do not provide a formal comparison with them. However, the off-the-shelf methods we consider in this paper can still be an alternative to complex approaches when further manual supervision is expected. More than acceptable recalls are obtained with huge reductions of the SNOMED CT term set. For example, if we consider just the top term retrieved by using the qGrams method along with Text piece of information a recall of 25.26% is obtained. This percentage rises up to 50.51% when considering the top ten terms, and to 75.56% if 100 terms are considered. That means a reduction of more than 99.99% on the SNOMED CT set size, which turns into an easy-to-handle set, yet obtaining a recall comparable to that of other approaches mentioned in the introduction. Statistical results have shown that at a level of confidence of 90%, the aforementioned TCM outperforms the rest of TCMs. Thus, its results may be used as a baseline against which results from sophisticatedly-designed new approaches could be compared, in order to determine whether, in each specific situation, it is worthy to use those complex methods or a simpler one is enough. Other conclusions can be derived by studying the performance of the TCMs considering their comparison method and their piece of information separately. On the one hand, the TCMs that use the qGrams method outperform or at least equal those using Jaccard similarity and Levenstein distance for each of the three pieces of information. Moreover, the TCMs that use Jaccard similarity outperform those using Levensteins distance for all three pieces of information. On the other hand, the TCMs that use the Text piece of information outperform those using Extended Texts or Descriptions for each of the three comparison methods, and the TCMs that use the Extended Texts outperform or at least equal those using the Descriptions for all three comparison methods. In summary, the syntactic-based methods alone are not strong enough to provide the correct bindings in all cases, but can be used to perform a huge reduction of the candidate set and later be complemented by other methods (e.g. semantic) or manual supervision. As future work we consider the use of SNOMED CT synonyms in our tests.

**Acknowledgments**

The authors would like to thank the authors in [12] for providing access to the manually coded archetypes.**References**

[1] Garde S, Knaup P, Hovenga E, Heard S. Towards Semantic Interoperability for Electronic Health Records. Methods of Information in Medicine, 2007;46: 332–343

[2] Berges I, Bermúdez J, Goñi A, Illarramendi A. Towards a satisfactory conversion of messages among agent-based information systems. Expert Systems with Applications. 2013; 40(7): 2462-2475

[3] Bird L, Goodchild A, Tun ZZ, Experiences with a Two-Level Modelling Approach to Electronic Health Records, Journal of Research and Practice in Information Technology. 2003; 35:(2)121–138.

[4] openEHR. 2013. Available from www.openehr.org

[5] Electronic Health Record Communication – Part 1: Reference Model. 2008

[6] HL7 CDA. 2013. Available from www.hl7.org/implement/standards/product_brief.cfm?product_id=7

[7] SNOMED CT. 2013. Available from www.ihtsdo.org/snomed-ct/

[8] Yu S, Berry D, Bisbal J. An investigation of semantic links to archetypes in an external clinical terminology through the construction of terminological "shadows". IADIS eHealth, 2010. Freiburg, Germany.

[9] Lezcano L, Sánchez S, Sicilia MA. Associating Clinical Archetypes through UMLS Metathesaurus Term Clusters. J. Medical Systems. 2012; 36(3): 1249-1258

[10] Unified Medical Language System. 2013. Available from www.nlm.nih.gov/research/umls/

[11] Qamar R. Semantic Mapping Of Clinical Model Data To Biomedical Terminologies To Facilitate Interoperability. 2008. University of Manchester.

[12] Meizoso M, Iglesias JL, Martínez D, Taboada MJ. Semantic similarity-based alignment between clinical archetypes and SNOMED CT: An application to observations. International Journal of Medical Informatics. 2012; 81(8): 566-578

[13] Available from http://www.usc.es/keam/TermArchetypes/input.html

[14] Levenshtein VI. Binary codes capable of correcting deletions, insertions and reversals. *Sov. Phys. Dokl.* 1966; 6:707-710

[15] Jaccard P. Etude comparative de la distribution florale dans une portion des Alpes et des Jura. Bulletin de la Société Vaudoise des Sciences Naturelles. 1901; 37(142):547-579.

[16] Implementation available from https://code.google.com/p/java-similarities/source/browse/trunk/simmetrics/src/main /java/uk/ac/shef/wit/simmetrics/similaritymetrics/QGramsDistance.java

[17] Friedman M. The use of ranks to avoid the assumption of normality implicit in the analysis of variance. J. Amer. Statist. Assoc. 1937; 32. 675-701.

[18] Demsar J. Statistical Comparisons of Classifiers over Multiple Data Sets. Journal of Machine Learning Research. 2006; 7: 1-30

[19] Garcia S, Herrera F. An Extension on "Statistical Comparisons of Classifiers over Multiple Data Sets" for all Pairwise Comparisons". Journal of Machine Learning Research. 2008; 9:2677-2694.